\documentclass[11pt,twoside,usenatbyb]{article}
\usepackage{asp2010}

\resetcounters

\begin{document}

\title{Initial parameters of neutron stars}
\author{S.B. Popov$^1$ and R. Turolla$^2$}
\affil{$^1$Sternberg Astronomical Institute,
Lomonosov Moscow State University, Russia}
\affil{$^2$Department of Physics and Astronomy, University of Padova,
Italy}

\begin{abstract}
A subpopulation of neutron stars (NSs), known as central compact objects
(CCOs) in
supernova remnants, are suspected to be low-field objects
basing on $P$~--~$\dot P$ measurements for three of them. The birth
rate of low-field NSs is probably
comparable with the birth rate of normal radio
pulsars. However, among compact objects in High-Mass X-ray Binaries (HMXBs)
we do not see robust candidates  for low-field NSs.
We propose that  this contradiction can be solved if magnetic fields of CCOs
was buried due to strong fall-back, and then the field emerges on the time
scale $10^4$~--$10^5$~yrs.
\end{abstract}

Our talk presented during the conference consisted of two parts.
In the first we presented results on estimates of initial spin periods
of neutron stars (NSs) associated with supernova remnants (SNRs). These
results are published already \citep{pt2012}, so in this contribution to the
conference proceeding we concentrate on the second part of the talk
related to a
possible evolution of one of subpopulations of isolated NSs.

Among young isolated NSs a very special group of objects is
recognized.
They appear as central compact objects (CCOs) in SNRs,
and some of them are dubbed ``antimagnetars'' \citep{hg2010}.
For these sources (three cases up to now)
periods about $P\sim$~0.1-0.5 s are measured,
together with period derivatives $\dot P\sim 10^{-17}$~s/s.
This allows to obtain an estimate of the magnetic field
$B\sim 10^{10}$~--~$10^{11}$~G.
It is suspected that all CCOs belong to the class of antimagnetars, and we
accept this hypothesis below.

More than 20 radiopulsars (PSRs) have robust associations with SNRs.
Average age of such systems is above $10^4$~yrs.
On the other hand, there are about 8 CCOs, and their ages are below
$10^4$~yrs, on average.  Estimates show that the
birthrate of antimagnetars is comparable with the birth rate of PSRs (even
taking into account beaming for PSRs).
Below we assume that about 1/3 of newborn NSs are antimagnetars.

Starting with these assumptions we can ask: what is the fate of
antimagnetars?
One possibility is that they just cool down and become invisible
as bright thermal X-ray sources.  Still, having 1/3 of NSs born with low
magnetic fields we can look at young X-ray binaries
to investigate further fate of such objects, assuming that the fraction of
low-field NSs is the same for isolated objects and (at least) for the
first-born NSs in binaries.
Let us discuss the population of
high-mass X-ray binaries (HMXBs).  These systems are young,
because the second massive companion is (usually) still on the main
sequence.
The majority of HMXBs belong to the class of Be/X-ray binaries.
Several other systems have wind-fed accretion.
This allows us to assume that total accreted
mass is not large enough to influence significantly the magnetic field of a
compact object, as it happened in millisecond PSRs.

A NS in a binary with $P_\mathrm{orb}\sim 10$~days with $B\sim
10^{10}$~G and accretion rate $\dot M\sim 10^{15}$~g/s is expected
to have $P\la 0.1$~s. This is valid for any regime of accretion:
standard disc, standard wind, or the new model proposed by
\cite{spkh2012}:

\begin{equation}
P_\mathrm{wind}=4.4 (P_\mathrm{orb}/10\,{\mathrm d})^{1/2} \mu_{30} \dot M_{16}^{-1/2}
(v/300 \,{\mathrm{km s}}^{-1})^2\, {\mathrm s}.
\label{wind}
\end{equation}

\begin{equation}
P_\mathrm{disc}=5.47 \mu_{30}^{6/7} \dot M_{16}^{-3/7} \, {\mathrm s}.
\label{disc}
\end{equation}

\begin{equation}
P_\mathrm{Shak}=8.18 \mu_{30}^{12/11} (P_\mathrm{orb}/10 \, {\mathrm d}) \dot M_{16}^{-4/11} (v/300 \,
{\mathrm{km}}{\mathrm  s}^{-1})^4 \, {\mathrm s}.
\label{shak}
\end{equation}

In catalogues of Be/X-ray binaries \citep{rp2005,
 reig2011}\footnote{http://xray.sai.msu.ru/~raguzova/BeXcat/}
 there are very few systems, in which NSs can have
low magnetic fields.
In the study by \cite{cp2012} it was shown that in a large (40 objects)
sample of Be/X-ray binaries in SMC there are
no NSs with fields below $10^{11}$~G.  Note, that many different methods to
derive field estimates have been used.

At this point we see a contradiction. About 1/3 of NSs in not-very-wide HMXBs
are expected
to have
periods below $\sim 0.1$ s, which corresponds to low magnetic field,
but very few such objects are observed.  Discussion of this contradiction is
the central point of this note.

Older CCOs are not known also among sources selected by their thermal
emission.  Among close-by cooling NSs only
normal pulsars and so-called Magnificent seven sources are known
\citep{baikal2011}.
However, even if CCOs follow slightly different cooling curves due to
accreted low-element envelops \citep{yak2004} they have to be bright at
least for $10^5$~yrs,
then older (than known CCOs) bright X-ray sources in SNR (or without them)
with CCO-like properties are expected.  There is some evidence that the
source
2XMM J104608.7-594306 can be a low-magnetized NS
(Mancini Pires et al. 2012\nocite{pires2012}, see also these proceedings),
but in this case the situation is not certain, yet, and there is a possibility
that spin period is short.
Population synthesis of close-by NSs does not require additional
contribution from low-field NSs \citep{popov2010}.
According to this study the global NS formation rate is consistent with a
single-mode field distribution,
in which there is no room for a large fraction of NSs with typical fields
below $10^{11}$~G.  A detailed discussion of
formation rates of NSs of different types was given by \cite{keane}.
However, these authors did not discuss CCOs (i.e., low-field NSs).
Still, inclusion of such objects will only strengthen their conclusion that
there shoud be evolutionary links between different
subpopulations of isolated NSs.

A possible solution can be related to emerging magnetic field in CCOs.
If after a strong fall-back episode NS's magnetic field was buried by a
thick layer of matter, then we can observe a
young hot NSs with low magnetic field \citep{ho2011,vp2012}.
Having $\ga 100$ HMXBs with typical lifetime $\sim 10^7$~yrs, we expect that
the youngest of them should have
ages $\sim10^5$~yrs.  Then, field should emerge on the timescale
few~$10^4$~--~$10^5$~yrs.


Now we have to look at normal PSRs. Emerging field effectively means that
there is ``injection'' of PSRs with $P\sim 0.1$--0.5 s
and normal magnetic field (on average, field can be slightly lower than the
standard value due to field decay, Ho 2011\nocite{ho2011}).
Surprisingly, this is what was discussed in many papers
\citep{vn1981,no1990}
starting from early 80s.  Later on, with new data these results were
reconsidered.

Still, \cite{vetal2004}
conclude that $\sim40$\% of pulsars can be
born with periods 0.1-0.5 s.
In a recent study \citep{pt2012}
we show that a significant fraction
of NSs can have such periods.
Among normal PSRs in SNRs this fraction (however, on low statistics)
is below 40\%.  Then, there is a room for pulsar ``injection'' which can
solve the problem of absence of low-field NSs in HMXBs.
%

%
Some contribution to PSRs with field $10^{10}$~--~$10^{11}$~G and periods
about few tens of second can be related to
disrupted binaries in which NSs have been mildly recycled \citep{belczynski}.
PSRs with long periods (for a given field and age) can appear due to fossil
discs \citep{yan2012}.

In a recent paper \cite{vm2011}
demonstrated that the
approach
based on ``pulsar current'' studies is very model dependent.
Then, results obtained by this methods should be taken with care.
We conclude, that the question of existence of pulsar ``injection''
with normal fields and $P\sim 0.1-0.5$~s is uncertain, but such a
possibility
due to emerging magnetic field on a time scale $\sim 10^5$~yrs
can solve the mystery of absence of significant number of low-field NSs in
HMXBs.


On $P$~--~$\dot P$ diagram emerged PSRs might appear at $P\sim 0.1$~--~0.5~s
and field slightly below the standard one.
Such PSR are expected to have negative braking indices, as their filed is
growing.
About 20-40 objects with such properties are known.
Growing magnetic field was also proposed for a young PSR J1734-3333
(Espinoza et al. 2011\nocite{espinoza}).
However, this is a PSR
with $P=1.2$~s and large magnetic field, i.e.  different from what we expect
for mature (emerged as PSRs) CCO-like NSs (see, however, Vigano \& Pons
2012\nocite{vp2012}).

Normal PSRs with detected thermal emission which does not fit well their
characteristic ages (too hot for a given age) can be former CCOs. In this
case their present day characteristic ages are not representative, and they
are still young to have high temperatures.
On other hand, NSs with buried (and then emerged) field should not be
necesserely hot, because their masses can be high enough to start rapid
cooling due to direct URCA processes. Then a NS becomes visible only after
field emergence on the PSR stage.


The only Be/X-ray binaries with very short periods are A 0538-66
and SAX J0635+0533. A0538-66 demonstrates episodes of very high
luminosity. Magnetic field for the NS is estimated as
$\sim10^{11}$~G.\footnote{The
 source SMC X-1 with $P=0.71$~s is also a bright object, and the field is
estimated to be normal: $\sim 10^{12}$~G.}
The situation with SAX J0635+0533 is more interesting as it has low
luminosity \citep{cus00}.
Before the paper by \cite{spkh2012} appeared
estimates of the magnetic field have produced very
low values.
Using eq.(\ref{disc}) or eq.(\ref{shak}) and standard assumptions we obtain
field $\la 10^{10}$~G for SAX J0635+0533.
Potentially, it can be an aged CCO-like NS with yet-non-emerged field.
This can point to the young age of a NS in this system.


X-ray pulsations can be hardly detectable if the spin axis is nearly
parallel to the magnetic dipole axis.
This is a possibility for low-field NSs to avoid identification via period
measurements in accreting systems, and so can be an alternative to field
emergence.
Note, that in the case of CCOs only in few cases periods are detected.
This, potentially, also can be linked to
small angles between spin and magnetic axis.  Among Be/X-ray binaries there
is a significant number -- roughly one third --
of sources with undetected periods \citep{rp2005}.
However, those objects
without spin period detection are mostly not well studied sources.
In addition, some of
them can be WD binaries
(like, most probably, $\gamma$ Cas, see Lopes de Oliveira et al. 2006\nocite{gcas}).
Significant alignment of spin and magnetic axis for low-field
sources on a time scale below one million years
is not expected in models of pulsar evolution, so we neglect this
possibility
here.
Note, that for HMXBs there is no evidence for alignment or counter-alignment
even for stronger magnetic fields \citep{ap2010}.
Then significant evolutionary alignment for low-field objects seems to be
not probable.

Another option can be related to smaller amplitude of pulsating signal for
low magnetic fields.
However, for not too high accretion rates in the case of well-studied
sources pulsations must be detectable even for low fields.


There is also an important
possibility that in close binaries with one episode of
strong mass
exchange  (all Be/X-ray binaries belong to this class)
fall-back is very small \citep{pejcha2012}. 
Potentially, this can result in smaller fraction of low-field NSs in
binaries.
 However note,
that to screen the field it is enough to have just a small
fall-back $\ll 0.1\, M_\odot$~\citep{ho2011,vp2012}.
Anyway, this option has to be
studied in more details.


We conclude that emerging magnetic field on the time scale
$\sim 10^4$~--~$10^5$~yrs
in CCO-like objects is the best possibility to explain the whole set of
data.


\acknowledgements
RT is
partially supported by ASI under a PRIN 2011 scheme. SBP is
also supported by RFBR grant 10-02-00059.
We thank participants of the conference  for
discussions and the Organizers  for this wonderful
meeting.

\bibliography{popov}

\end{document}